\documentclass[preprint2]{aastex63}

\usepackage{comment}
\usepackage{changepage}

\newcommand{\SolarMass}{$\mathrm{M_\odot}$}
\DeclareRobustCommand{\orderof}{\ensuremath{\mathcal{O}}}

\newcommand{\afflone}{Australian Research Council Centre of Excellence for Gravitational Wave Discovery (OzGrav) }
\newcommand{\affltwo}{Department of Physics, University of Western Australia, Crawley WA 6009, Australia}
\newcommand{\afflthree}{Department of Computer Science and Software Engineering, University of Western Australia, Crawley WA 6009, Australia}

\shorttitle{SPIIR Early Warning}
\shortauthors{Kovalam et al.}

\begin{document}
\title{Early Warnings of Binary Neutron Star Coalescence using the SPIIR Search}

\author[0000-0001-8143-9696]{Manoj Kovalam}
\email{manoj.kovalam@research.uwa.edu.au}
\affiliation{\afflone}
\affiliation{\affltwo}

\author[0000-0003-0760-3835]{Md Anwarul Kaium Patwary}
\affiliation{\afflone}
\affiliation{\afflthree}

\author{Anala K. Sreekumar}
\affiliation{\afflone}
\affiliation{\affltwo}

\author[0000-0001-7987-295X]{Linqing Wen}
\email{linqing.wen@uwa.edu.au}
\affiliation{\afflone}
\affiliation{\affltwo}

\author[0000-0002-2618-5627]{Fiona H. Panther}
\affiliation{\afflone}
\affiliation{\affltwo}

\author{Qi Chu}
\affiliation{\afflone}
\affiliation{\affltwo}

\begin{abstract}

Gravitational waves from binary neutron star mergers can be used as alerts to enable prompt follow-up observations. In particular, capturing prompt electromagnetic and astroparticle emissions from the moment of a binary merger presents unique constraints on the time scale and sky localization for online gravitational wave detection. Here we present the expected performance of the SPIIR online detection pipeline that is designed for this purpose in the upcoming international LIGO-Virgo's 4th Science Run (O4). Using simulated Gaussian data for the two LIGO observatories with expected O4 sensitivity, we demonstrate that there is a non-negligible opportunity to deliver pre-merger warnings at least $10\,\mathrm{s}$ before the final plunge. These alerts are expected to be issued at a nominal rate of one binary neutron star coalescence per year and localized within a median searched area of $300\,\mathrm{deg^2}$. We envision such a detection to be extremely useful for follow-up observatories with a large field of view such as the Murchison Widefield Array radio facility in Western Australia.

\end{abstract}

\keywords{Gravitational wave astronomy (675), Compact binary stars (283), Gravitational wave sources (677), Neutron stars (1108)}

\section{Introduction}

The era of gravitational wave (GW) astronomy began with its first ever detection from a compact binary coalescence (CBC) in 2015 \citep{LIGOScientific:2016aoc}, during the first advanced LIGO observing run. This was achieved using the two LIGO observatories at Hanford (H1) and Livingston (L1), \citep{LIGOScientific:2014pky}. In 2017, Virgo (V1, \cite{VIRGO:2014yos}) joined the duo in the second observing run. The iconic detection of a binary neutron star (BNS) merger \citep{LIGOScientific:2017vwq} and associated electromagnetic (EM) emission signaled the beginning of a new era of GW multi-messenger astronomy (MMA). This was a joint detection of GWs and a short gamma-ray burst (sGRB) detected $\sim2\,\mathrm{s}$ after the binary merger, by the Fermi-GBM and \textit{INTEGRAL} space telescopes \citep{LIGOScientific:2017ync}. This event was followed by numerous other EM observations spanning the entire EM spectrum. This detection has had a tremendous impact in astronomy with several studies emerging from it, including estimation of the Hubble constant \citep{LIGOScientific:2017adf}, constraints on the neutron star equation of state \citep{LIGOScientific:2018hze}, and also connecting sGRBs and kilonovae to BNS mergers \citep{LIGOScientific:2017ync}.

GW detections were reported publicly in real-time for the first time during the third LIGO-Virgo observing run O3 \citep{aLIGO:2020wna, Virgo:2019juy}, leading to the search for EM counterparts of GW observations. However, the typical delay between a GW detection and the associated GCN alert was on the order of at least minutes \citep{Magee:2021xdx}, and hence the discovery of prompt EM emissions, which are expected during a binary merger \citep{Rezzolla:2011da}, continued to rely on serendipitous discovery. For example, in the case of GW170817, the sGRB was observed only two seconds after the GW detection. Thus, an advance warning of such events is crucial to alert conventional EM telescopes for prompt follow up observations.

Five gravitational wave detection pipelines have processed GW data in real-time for the past LIGO-Virgo-Kagra collaboration (LVK) science runs. Four modelled search pipelines - SPIIR \citep{Luan:2011qx, Hooper:2011rb, Chu:2020pjv}, GstLAL \citep{Messick:2016aqy, Sachdev:2019vvd}, PyCBC \citep{Nitz:2018rgo, DalCanton:2020vpm} and MBTA \citep{Adams:2015ulm, Aubin:2020goo} - use known CBC waveforms to identify a signal within the detector data, while burst signals are recovered by the cWB \citep{Klimenko:2015ypf} pipeline via a coherent analysis. SPIIR (Summed Parallel Infinite Impulse Response) uses a time-domain filtering method equivalent to matched filtering to detect GWs. The SPIIR pipeline uses GPU acceleration for parallel processing to reduce latency and improve computational efficiency \citep{Liu:2012vw, Guo:2018tzs}. Out of the 38 detections reported by SPIIR through public alerts in O3 \citep{Chu:2020pjv}, $90\%\,(35)$ of them were confirmed by the offline searches and later added to the GW catalogs \citep{LIGOScientific:2020ibl, LIGOScientific:2021usb, LIGOScientific:2021djp}. The overall latency for the SPIIR pipeline, defined as the time delay between a GW event merger and its detection, is $\sim9\,\mathrm{s}$,\footnote{not encompassing latencies associated with data transfer from detectors and skymap/alert generation\label{note1}} which corresponds to the H1L1 two detector analysis. This paper describes a new feature for the SPIIR search, targeting 'early warning' to detect GWs and generate alerts before the merger or at negative latencies. Note that a binary coalescence involving at least one neutron star, tends to spend a few minutes of its time during the inspiral phase within the LIGO-Virgo sensitive frequency band $(15-2000\,\mathrm{Hz})$. 

Early warning alert systems associated with other matched filtering pipelines have been tested for their localization accuracy and detection efficiency\textsuperscript{\ref{note1}} \citep{Sachdev:2020lfd, Nitz:2020vym}. In 2020, a mock data challenge \citep{Magee:2021xdx} was conducted within the LVK to test the capability of the low-latency infrastructure to send EW alerts. SPIIR participated in this test sending out one of the five EW mock GCNs and successfully demonstrated the feasibility for it to send pre-merger alerts.\footnote{https://gcn.gsfc.nasa.gov/gcn3/27989.gcn3} This study also estimated the expected rate of BNS mergers and their localization areas using \texttt{Bayestar} rapid localization \citep{Singer:2015ema}, by simulating a four-detector network H1, L1, V1 and Kagra (K1, \cite{KAGRA:2020tym}) in O4.

In this work, we analyse the performance of the SPIIR EW pipeline in a simulated O4 environment. We, in particular, study the performance of the two LIGO detector network (H1L1) which has the minimum expected overall latency caused by data transfer among all available detectors. The main motivation behind the choice of this network is to make detection as early as possible, thereby assisting EM telescopes to observe possible short transients $(\sim{1}\,\mathrm{s})$ right at the merger \citep{Rezzolla:2011da}. Since most follow-up telescopes need time to orient themselves to the source direction, saving an additional $4\,\mathrm{s}$ of the delay from Virgo is extremely useful. We wish to take advantage of this faster network even though it results in a possibly worse localization, which can still be beneficial to observatories with large fields of view (FOV). It should be noted that Virgo will be added to the full-bandwidth search which will produce an enhanced localization within tens of seconds. We then demonstrate the accuracy of parameters including signal-to-noise ratio (SNR) and the chirp mass, which are estimated internally and used to classify the source type via p\_astro \citep{Kapadia:2019uut} and infer properties via hasNS \citep{Foucart:2018rjc}. We also ensure the reliability of localization areas at various latencies, which are published via SPIIR EW alerts.  

This paper is organized as follows. In section~\ref{sec:method}, we show details about the SPIIR EW method, and provide information about the simulated O4 data and GW signal injections. Section~\ref{sec:results} discusses the results of this simulation run. Finally, in section~\ref{sec:disc}, we discuss the results in the context of EM follow-up by ground and space-based facilities.

\section{Method}\label{sec:method}

Early warning alerts are expected to be issued publicly for the first time by LVK in O4. Through this study we demonstrate the expected performance of the SPIIR online pipeline to deliver EW alerts. We only use the LIGO detectors in this work because this two detector network has a very high coincident duty-cycle time ($62\%$ of total) reported in O3 \citep{LIGO:2021ppb}, high sensitivity with nearly aligned antenna beam patterns, and most importantly, saves an additional four second data transfer latency caused by Virgo \citep{Magee:2021xdx}, making it the most viable combination to promptly report a pre-merger alert in O4, at the cost of a poorer localization for some of the bright events at early detections as compared to other combinations.

\subsection{SPIIR Early Warning pipeline}

Matched filtering is the optimal method used by CBC pipelines to detect the presence of a GW signal amidst noise. This method involves cross correlating known waveforms, also known as templates, with detector data to output SNR \citep{Finn:1992wt, Cutler:1994ys}. SPIIR filtering is a time-domain equivalent to matched filtering \citep{Luan:2011qx, Hooper:2011rb} which uses first order IIR filters to approximate GW templates to a high accuracy, which are then used in a time-domain convolution with the detector data, constructing the SPIIR SNR. The best matched template would maximize the SNR in the presence of a signal, assuming that the noise is stationary Gaussian.

SPIIR pipeline has the capability to process GW data in extremely low latency. Since it is a time-domain convolution, it takes SPIIR theoretically close to $0\,\mathrm{s}$ to produce SNRs (\cite{Luan:2011qx,Hooper:2011rb}). However, in our actual implementation, the design costs SPIIR $\sim1\,\mathrm{s}$ to compute the SNR and $\sim4\,\mathrm{s}$ to identify candidates \citep{Chu:2020pjv}. A coherent network SNR across all active detectors is calculated for any single-detector event above a particular SNR threshold (set to $\rho > 4$ for real-time searches). Trigger candidates are then ranked based on two quantities: (a) the reduced chi-square ($\xi^2$), indicating a goodness-of-fit of the SNR time series against expectation and (b) the coherent network SNR ($\rho_\mathrm{c}$). A false alarm rate (FAR) is calculated based on the probability of an event associated with the background.

In this work, we use a template bank consisting of $100,000$ binary mass and projected spin pairs. These waveform templates have component masses ranging from 1.1\,\SolarMass\,\textless\,$\mathrm{m_1,m_2}$\,\textless\,3.0\,\SolarMass, focusing on binaries with two neutron stars with promising opportunities to observe EM emissions \citep{Rezzolla:2011da}. Waveforms for these masses (based on the SpinTaylorT4 \citep{Buonanno:2009zt} time domain approximation) are truncated at specific time intervals before merger and used to construct the SPIIR EW filters. The SPIIR filters used in the work are set to have a high ($>97\%$) overlap, which is the inner product of the original waveform and the approximant waveform.

We conduct $7$ parallel EW searches on the simulated data using the SPIIR pipeline, with the searches having their templates truncated between $10\,\mathrm{s} - 70\,\mathrm{s}$, at $10$ second intervals respectively. For simplicity, we label these configurations with their pre-merger truncation time. For example, the EW search with its templates truncated $10\,\mathrm{s}$ pre-merger is simply addressed as the $-10\,\mathrm{s}$ run. It should be noted that this number only represents the pre-merger template truncation time and does not correspond to the overall detection latency, which will be addressed in section~\ref{sec:latency}. The EW searches are processed at a sampling rate of $256\,\mathrm{Hz}$ in conjunction to a full-bandwidth simulation (non-truncated templates), which is processed at $2048\,\mathrm{Hz}$ for comparison. This is because EW detections are recovered at a frequency $< 128$ Hz at $10\,\mathrm{s}$ before the merger and earlier, in the inspiral phase.\footnote{The downsampling also reduces the computational complexity of EW simulations, thereby decreasing the amount of resources used.} We label the full-bandwidth simulation as the $0\,\mathrm{s}$ run to stay consistent across. For each of these searches, the FARs associated with candidate triggers are computed independent from one another.

\subsection{Simulated data}\label{sec:simdata}

We inject signals into stationary Gaussian noise simulating the strain data produced by the LIGO detectors in O4. The estimated power spectral density (PSD) of LIGO in O4 is given in \cite{KAGRA:2013rdx}, with an expected BNS range of $190\,\mathrm{Mpc}$. We used the \texttt{gstlal\_fake\_frames} package \citep{Messick:2016aqy} to generate this strain data using the given PSD. Both LIGO-Hanford and LIGO-Livingston are expected to reach a similar sensitivity in O4 and hence the output strain is expected to be similar in both. The whole data segment spans three weeks and is generated in the \texttt{gwf}\footnote{https://dcc.ligo.org/LIGO-T970130/public} format.

The injected signals used in this analysis are generated using the \texttt{lalsuite} package \citep{lalsuite}. The component masses of the injections are sampled between 1.0\,\SolarMass\,\textless\,$\mathrm{m_1,m_2}$\,\textless\,2.3\,\SolarMass, with a uniform distribution. The spin is restricted to be below $0.4$ for both components and sampled with an isotropic distribution. The injected signals are distributed uniformly within a comoving volume of redshift $z=0.2$. The final population set, using the constraints mentioned above, has \orderof{($10^4$)} number of injections.

\begin{figure}[tb!]
    \centering
	\includegraphics[width=\linewidth,trim={0.7cm 0cm 2cm 1.5cm},clip]{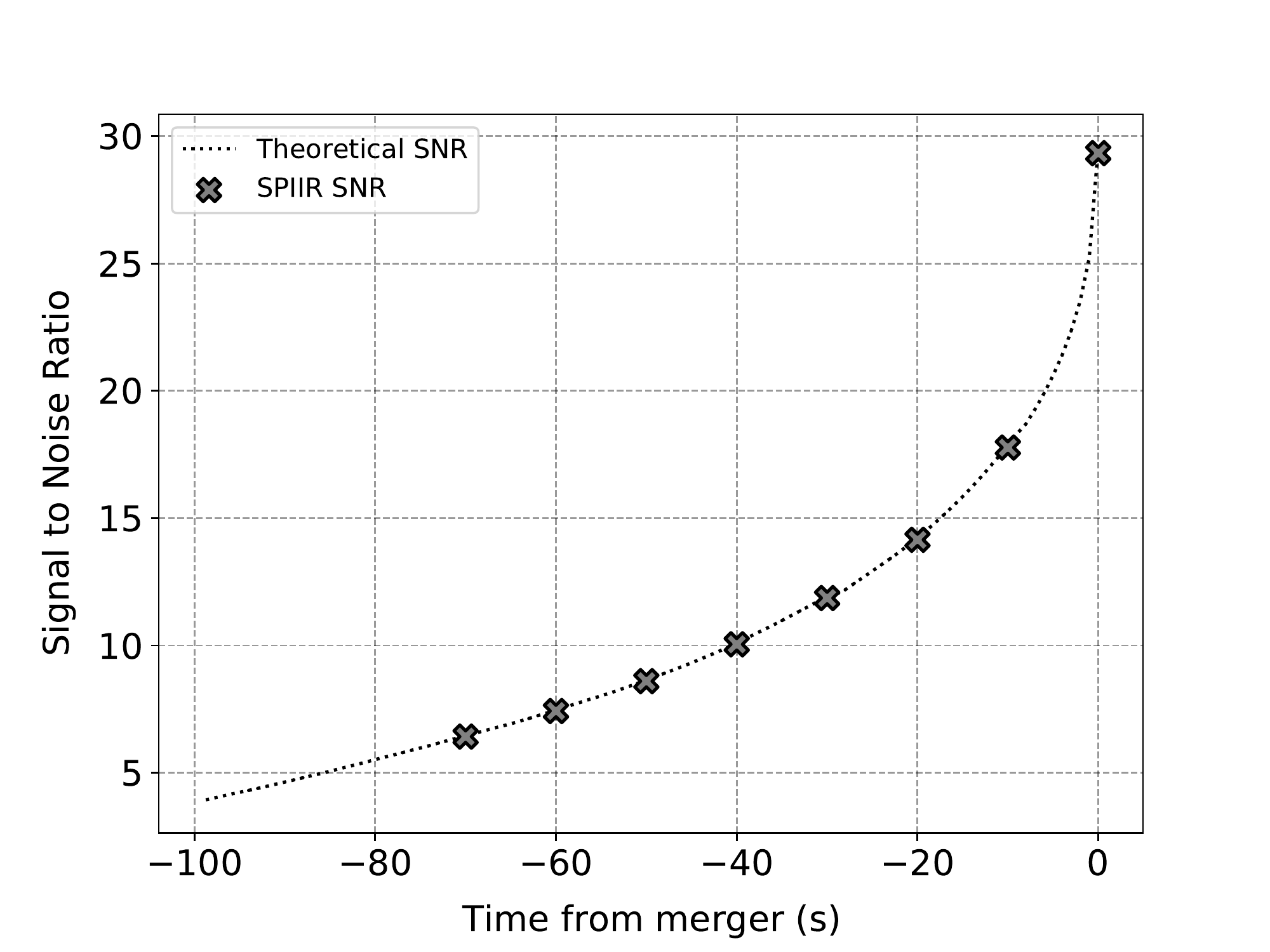}
	\caption{Evolution of SNRs at different pre-merger latencies in simulated LIGO O4 data for a $1.4$\SolarMass$-1.4$\SolarMass BNS system at a distance of $100$ Mpc. The dotted line represents the theoretical SNRs calculated using matched filtering while the symbols show SNRs from the SPIIR method. A strong event with a full-bandwidth SNR of $30$ can be detected at sub-minute pre-merger latencies.}
	\label{fig:snrevo}
\end{figure}

 Expected SNRs for GW signals can be theoretically calculated using their mass, spin and distance parameters. By truncating the waveforms, we can also estimate the SNRs at different latencies before the merger. Fig.~\ref{fig:snrevo} shows the evolution of SNRs at different pre-merger latencies for a fiducial BNS reference $1.4\,\mathrm{M_\odot}+1.4\,\mathrm{M_\odot}$ source at $100\,\mathrm{Mpc}$ in O4 sensitivity. We also show that the SPIIR SNRs at the truncated intervals indicate consistency with the expected SNRs. Similarly, expected SNRs at different pre-merger latencies are theoretically calculated for all the injected signals in our simulation.
 
\section{Results}\label{sec:results}

Recovered signals from the searches are considered to be a detection only if the FAR reported by the associated search is less than one per month (or $3.85\times10^{-7}\,\mathrm{Hz}$), which is also the threshold used for reporting EW candidates in the MDC \citep{Magee:2021xdx}. We find that $61$\% of all injected signals were detected by the $0\,\mathrm{s}$ run. Using $320^{+490}_{-240}\, \mathrm{Gpc}^{-3}\text{yr}^{-1}$ as the local BNS merger rate, estimated in \cite{LIGOScientific:2020ibl}, we expect a median detection rate of $\sim18$ BNSs per year in O4 using the full-bandwidth configuration. It should be noted that our choice of the FAR threshold also includes sub-threshold events, when compared to the threshold of one per two months used to select the open public alerts during O3. In comparison, \cite{Magee:2021xdx} uses a threshold of $\mathrm{SNR}>12$ and anticipate the detection rate to be $\sim9$ BNSs per year for a H1L1V1K1 four detector network. If we apply the same criterion, we also see a similar rate but this is for the H1L1 two detector network. This is probably due to the fact that \cite{Magee:2021xdx} uses a Gaussian distribution for their mass model, while we use a uniform distribution, which leads to a larger detection range for our search and hence a comparatively larger rate.

\begin{figure}[t!]
    \centering
	\includegraphics[width=\linewidth,trim={0.45cm 0.4cm 0.45cm 0.1cm},clip]{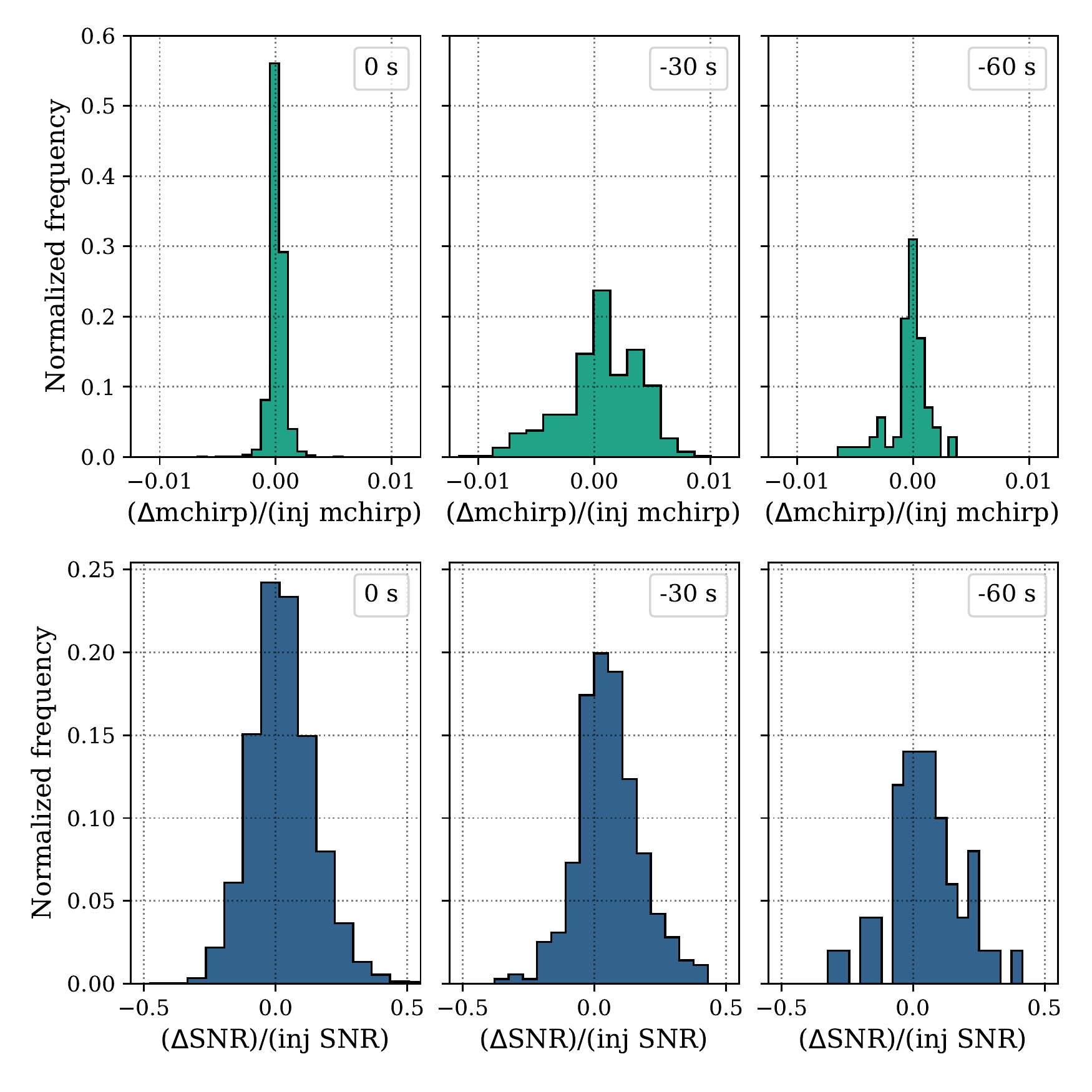}
	\caption{Accuracy of chirp mass and SNR recovery for the $0\,$s, $-30\,$s and $-60\,$s configurations. Top panel: Distributions for fractional differences in recovered and injected chirp masses. The error margin is within $2\%$ for all the runs. Bottom panel: Distributions for fractional differences in recovered and expected SNRs. About half of all events fall within $10\%$ error margin (in the bottom panel), of which $3\%$ can be attributed to the SPIIR GW waveform approximation.}
	\label{fig:snr-mchirp}
\end{figure}

\subsection{Accuracy}\label{sec:accuracy}

Online estimations of parameters like chirp mass and SNRs are calculated internally by the pipeline and not published directly to the public. However, these parameters are essential to compute the published values of FARs, p\_astro \citep{Kapadia:2019uut} and hasNS \citep{Foucart:2018rjc}, which are used to infer source properties and classify detections. Thus, it is very important for pipelines to determine the accuracy of these parameters to ensure the reliability of the published alerts. Fig.~\ref{fig:snr-mchirp} compares the SNRs and chirp masses recovered from the searches to the injection set. We demonstrate the performance of the SPIIR EW pipelines by first comparing the chirp mass of the recovered signals to the expected values (Fig.~\ref{fig:snr-mchirp}, upper panel).
We find the fractional difference between the two to be less than $2\%$ for all runs, indicating an accurate recovery. We see that the chirp mass recovered from the $0\,\mathrm{s}$ run is within an error margin of $0.5\%$, with subsequently wider margins for other EW configuration. Next we find that, at an average, the recovered SNRs are within an error margin of $\sim10\%$ with the expected SNRs (Fig.~\ref{fig:snr-mchirp}, lower panel). It should be noted that a $3\%$ deviation is expected in SNRs because of the SPIIR waveform approximation. The rest of the discrepancy could be attributed to factors like the discreteness in template banks leading to a mismatch with injections, noise influences in low-SNR events and relatively minor PSD mismatches while calculating the SNRs using moving PSD estimations. The accuracy of chirp mass and SNR estimates for the full-bandwidth run has been well studied in \cite{Chu:2020pjv} with similar results reported. Fig. \ref{fig:snr-mchirp} shows a histogram of the fractional differences in injected and recovered parameters for $0\,\mathrm{s}$, $30\,\mathrm{s}$ and $60\,\mathrm{s}$ runs.\footnote{$-60$~s simulation has a lower number of statistics as compared to others.\label{note2}}

\begin{figure}[t!]
    \centering
    \vspace*{0.25cm}
	\includegraphics[width=\linewidth,trim={0.25cm 0cm 0.4cm 0.25cm}]{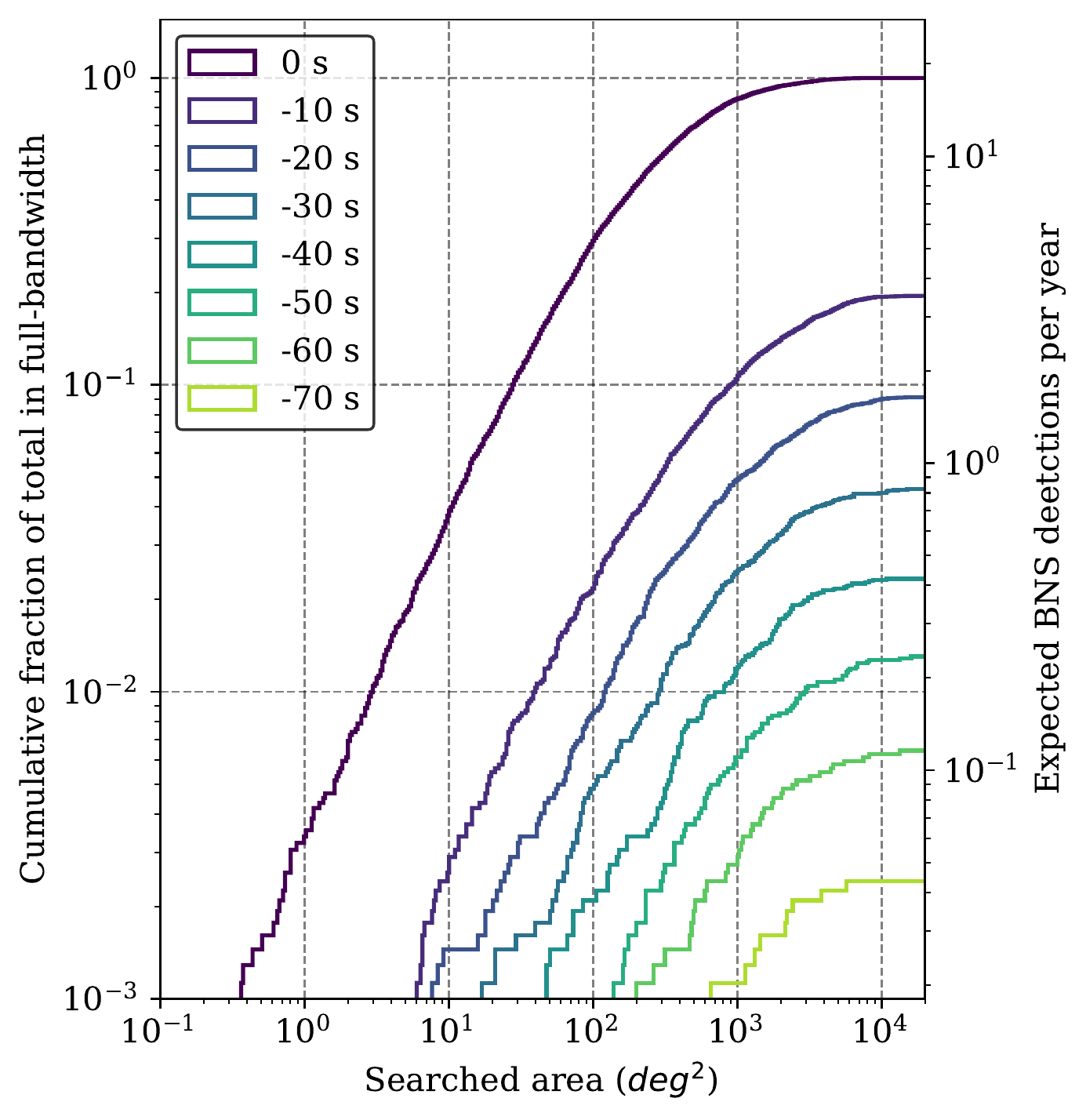}
	\caption{Searched areas vs number of detected events in each EW configuration as a cumulative fraction of the total number of full bandwidth detection (left y-axis). The secondary y-axis shows the same as function of expected number of BNS detections per year in O4. Each color represents one EW configuration. Using a simulated Gaussian environment for O4, we expect to recover at least $1(6\%)$ pre-merger event per year with a localization area $< 300\,\mathrm{deg^2}$ with the H1L1 configuration in O4.}
	\label{fig:localization}
\end{figure}

\subsection{Localization}\label{sec:loc}

Accurate and rapid localization of GW signals is crucial to enable prompt followup observations. Providing source direction to EM observatories via pre-merger alerts would help enable a prompt follow-up observation. For this study, we use LVK’s rapid localization software, \texttt{Bayestar} \citep{Singer:2015ema}, to construct sky maps. As we go down in latency, EW simulations trigger at lower frequencies, so only a partial information of the detected signal is recovered. This lowers the estimated SNRs and also the accuracy with which the temporal and phase information is recovered, thus resulting in larger localization areas \citep{Fairhurst:2017mvj}. Fig.~\ref{fig:localization} shows the cumulative histogram of the searched areas (defined as the smallest area needed to be searched to encompass an event) normalized w.r.t. the total number of detection in the full-bandwidth run ($18$). 

\begin{figure}[t!]
    \centering
	\includegraphics[width=\linewidth,trim={0.25cm 0cm 0.2cm 0cm},clip]{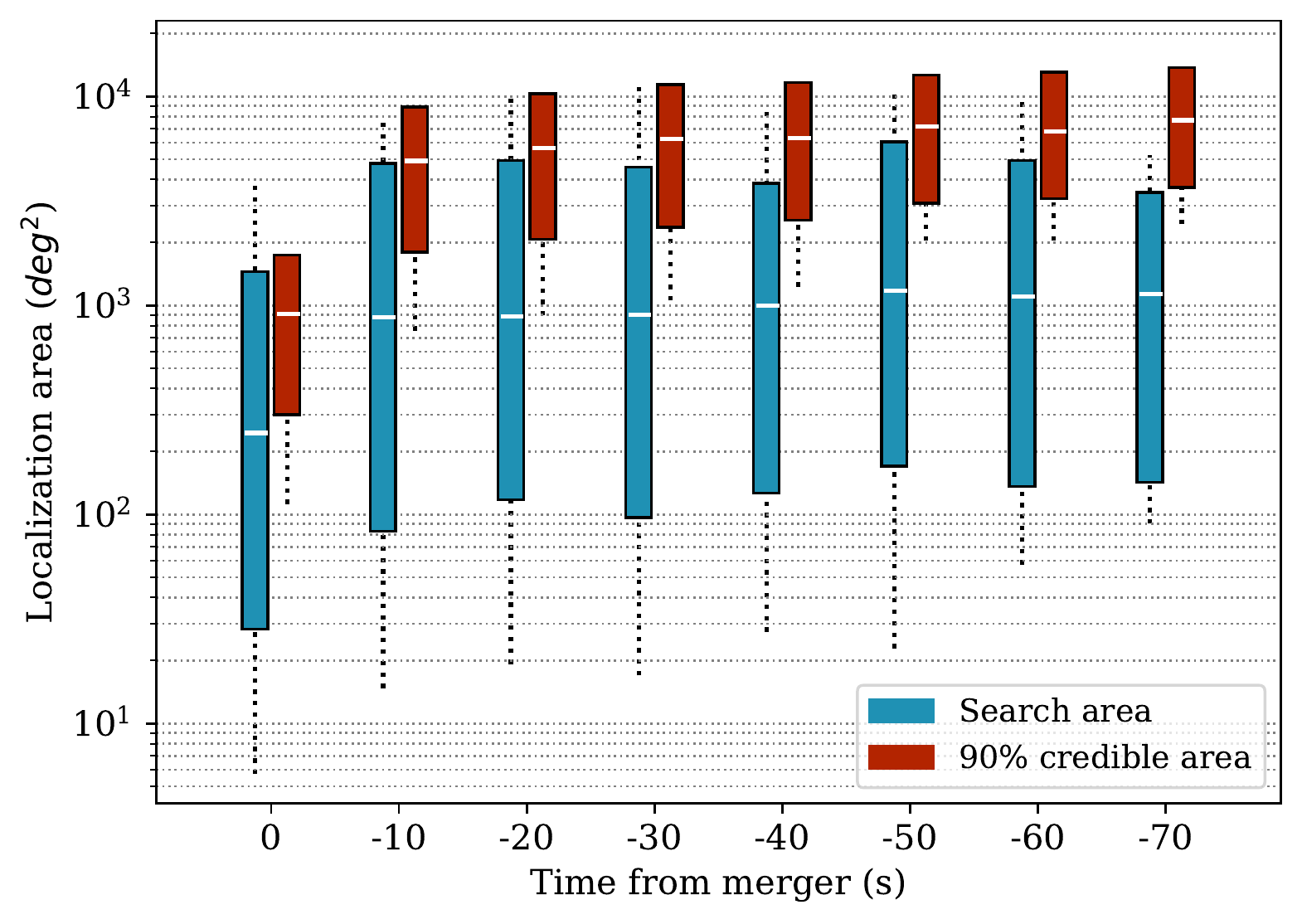}
	\caption{Box and whiskers plot comparing searched and $90\%$ credible areas for all the runs. The y-axis represents the localization areas and the x-axis represents the truncation latency for each EW run. The boxes encompass $95\%$ of the events and the whiskers extend up to the rest. The white lines within the boxes represent the median values of the respective data sets. Using simulated Gaussian data, we demonstrate that the median searched areas are several times lower than the $90\%$ credible areas, implying that a lot less area (as compared to the $90\%$ credible region) can be searched to localize the GW event in O4.}
	\label{fig:box}
\end{figure}

Based on our Gaussian noise simulation for O4, we expect to deliver pre-merger BNS alerts at a rate of $1\,(2)$ per year in O4 with a searched area below $300\,(1000)\,\mathrm{deg^2}$ using the LIGO network alone. We also expect to detect at least one pre-merger detection in two years localized within a searched area of $100\,\mathrm{deg^2}$ in O4. However, the chance to have a sub-minute pre-merger detection is seen to be less than $0.2\%$. We also compare the searched areas of the candidates with their $90\%$ credible region areas in Fig. \ref{fig:box}. Table \ref{table:areas} shows the median values for the searched and $90\%$ credible areas, and also the expected detection rate at all EW latencies in O4. We find that the median values of searched areas are several times smaller than $90\%$ credible areas and the difference becomes more prominent as we go down in latency for the EW runs.

\begin{table}[t!]
 \begin{adjustwidth}{-1.45cm}{}
 \resizebox{1.25\columnwidth}{!}{
  \begin{tabular}{c c c c c}
 \hline
 $\mathrm{EW}$ & $\mathrm{N_{events} (yr^{-1})}$ & $\mathrm{Area_{se} (\text{deg}^{2})}$ & $\mathrm{Area_{90} (\text{deg}^{2})}$ & $\mathrm{Latency (s)}$ \\
 \hline\hline 
 \vspace{.05in}
 $0\,$s & $18$ & $245$ & $911$ & $9.46^{+1.42}_{-0.76}$ \\
 \vspace{.05in}
 $-10\,$s & $3.51$ & $876$ & $4912$ & $-2.30^{+1.15}_{-1.20}$ \\
 \vspace{.05in}
 $-20\,$s & $1.64$ & $886$ & $5673$ & $-12.36^{+1.31}_{-1.15}$ \\
 \vspace{.05in}
 $-30\,$s & $0.83$ & $902$ & $6243$ & $-22.35^{+1.33}_{-1.18}$ \\
 \vspace{.05in}
 $-40\,$s & $0.42$ & $995$ & $6299$ & $-32.32^{+1.29}_{-1.17}$ \\
 \vspace{.05in}
 $-50\,$s & $0.23$ & $1178$ & $7187$ & $-42.27^{+1.19}_{-1.18}$ \\
 \vspace{.05in}
 $-60\,$s & $0.12$ & $1104$ & $6803$ & $-52.25^{+1.13}_{-1.14}$ \\
 
 $-70\,$s & $0.04$ & $1136$ & $7683$ & $-61.86^{+0.81}_{-0.82}$ \\ [1ex] 
 \hline
  \end{tabular}
}
 \end{adjustwidth}
\caption{Expected BNS detection rate ($\mathrm{N_{events}}$), median searched area ($\mathrm{Area_{se}}$), median $90\%$ credible area ($\mathrm{Area_{90}}$) and the median latency (with $90\%$ error margins) associated with different EW simulations for O4.}
\label{table:areas}
\end{table}

\subsection{Latency}\label{sec:latency}

The overall pipeline latency for each configuration is measured on an online data streaming platform and the median values are recorded in Table \ref{table:areas}. These values correspond to the internal latency of the pipeline, which includes delays in the SPIIR filtering algorithm and calculating the FARs. We find that the full-bandwidth configuration has a median intrinsic latency of $9-10\,\mathrm{s},$\footnote{with the potential to go below $5s$.} consistent with results from \cite{Chu:2020pjv}. Additional latencies including data-transfer, localization and alert-generation depend on the online infrastructure, which are believed to be significantly lower in O4 as compared to O3~\citep{Magee:2021xdx}. The EW runs have a negative median latency indicating pre-merger detection.

\section{Discussion}\label{sec:disc}

In this work, we investigated the prospects of using the SPIIR EW search, on the two-detector LIGO network, to detect GWs from binary neutron stars before their merger in O4. This configuration saves an additional $4\,\mathrm{s}$ of data transfer latency from Virgo making it the most probable combination to issue negative latency alerts in O4. However, the major drawback of using this network is that the source localization areas for bright events are poorer as compared to three or four detector networks. This is evident from the full-bandwidth detection of \texttt{GW170817} during O2, where the localization area improved by a factor of $\sim{6}$ with the inclusion of an additional detector \citep{LIGOScientific:2017vwq}. The main motivation behind our choice of a two detector network in O4 is to facilitate the rapid follow-up detection of short transient signals (only $\sim{1}\,\mathrm{s}$ long) within seconds of a BNS merger \citep{Rezzolla:2011da} that was not possible before. Given the following facts that: (a) most major follow-up telescopes require a response time of $\textgreater\,\sim{10}\,\mathrm{s}$ to reorient themselves to a particular sky direction, (b) the anticipated event rate is small for an early detection of BNS events in O4 (even just $10\,\mathrm{s}$ before their final merger), and (c) the latencies of the existing pipelines are larger than $10\,\mathrm{s}$, it is extremely challenging to observe short transients right at the merger. Thus, saving a few seconds of latency might prove to be extremely useful to capture such events using instruments with larger viewing areas and fast response times. Configurations including other detectors can also be run in parallel for better localizations but with a slight delay.

Using a Gaussian noise simulation, we show that EW pipelines have an exceptional recovery of chirp mass, which is helpful in classifying the source properties such as p\_astro and embright. We also compare the $90\%$ credible areas and the searched areas associated with the EW triggers, and demonstrate that the actual searched area for a detection is about one order of magnitude better. These areas are still a few times larger than the three and four detector network localizations reported in \cite{Magee:2021xdx, Sachdev:2020lfd, Nitz:2020vym}, but this trade-off to several to tens of seconds in latency is something we hope to take in O4. It is important to note that a more accurate three-detector and four-detector localization will be provided by the full-bandwidth SPIIR search.

We demonstrate the possibility of issuing alerts for atleast one BNS merger per year $\sim 12\, \mathrm{s}$ before the merger in table~\ref{table:areas}. Adding an additional $8\,\mathrm{s}$ for data transfer and localization, this would still be a prospect for a negative latency alert. The best localized EW alert is likely to be from the $-10\,\mathrm{s}$ search, and we expect to deliver alerts localized within $300\,\mathrm{deg^2}$ at a nominal rate of one detection per year from this search.  

These alerts are especially useful for follow-up observatories which have a large FOVs. Radio telescopes like the Murchison Widefield Arrray (MWA, \cite{Tingay:2012ps}) is one such benefactor, which has the capability of viewing a quarter of the sky. \cite{James:2019xca} has investigated the response time of this MWA observational mode to be about $10\,\mathrm{s}$. In the future, EW alerts produced by SPIIR can also be ingested by x-ray and gamma-ray missions with large FoVs. For example, the Neil Geherels Swift Observatory (FoV $\sim 4600\,\mathrm{sq.\,degrees}$) has developed a fully-autonomous, extremely low-latency on-board commanding pipeline (GUANO, \cite{Tohuvavohu:2020stm}) capable of recovering sub-threshold Burst Alert Telescope (BAT) triggers. Similarly, the large fields of view of the \textit{FERMI} gamma-ray telescope's gamma-ray burst monitor (FoV $\sim 3.2E4\,\mathrm{sq.\,degrees}$, \cite{Meegan:2009qu}) and \textit{INTEGRAL}'s gamma-ray burst detection sub-systems (FoV all-sky for SPI-ACS, except for regions occulted by the Earth, \cite{RauSPIACS}) are also well suited to following up SPIIR triggers. Thus, the SPIIR EW pipeline will contribute significantly to the possibility for targeted followup of sGRB signals by the global astronomy community.

\section{Acknowledgements}

This research was supported by the University of Western Australia and funded by the Australian Research Council (ARC) Centre of Excellence for Gravitational Wave Discovery OzGrav under grant CE170100004. This work relied on the computational resources provided by the LIGO Laboratory at California Institute of Technology and the OzStar super-computing cluster at Swinburne University of Technology. The authors are grateful to them. The LIGO Laboratory cluster is funded by National Science Foundation Grants PHY-0757058 and PHY-0823459 and the OzStar program receives funding in part from the Astronomy National Collaborative Research Infrastructure Strategy (NCRIS) allocation provided by the Australian Government. We thank Jarrod Hurley and Stuart Anderson for the resources provided on the computing clusters and acknowledge the efforts by Alex Codoreanu and Patrick Clearwater in maintaining the SPIIR repository.

\bibliography{spiir-ew.bib}{}
\bibliographystyle{aasjournal}

\end{document}